# A New Kind of Economy is Born
## Social Decision-Makers Beat the "Homo Economicus"

by Dirk Helbing (ETH Zurich)

**The Internet and Social Media change our way of decision-making. We are no longer the independent decision makers we used to be. Instead, we have become networked minds, social decision-makers, more than ever before. This has several fundamental implications. First of all, our economic theories must change, and second, our economic institutions must be adapted to support the social decision-maker, the "homo socialis", rather than tailored to the perfect egoist, known as "homo economicus".**

The financial, economic and public debt crisis has seriously damaged our trust in mainstream economic theory. Can it really offer an adequate description of economic reality? Laboratory experiments keep questioning one of the main pillars of economic theory, the "homo economicus". They show that the perfectly self-regarding decision-maker is not the rule, but rather the exception [1,2]. And they show that markets, as they are organized today, are undermining ethical behavior [3].

Latest scientific results have shown that a "homo socialis" with other-regarding preferences will eventually result from the merciless forces of evolution, even if people optimize their utility, when offspring tend to stay close to their parents [4]. [1] Another, independent study was recently

---

[1] Experts should note that there has been research on so-called "altruistic behavior" in social dilemma situations such as the prisoner's dilemma since more than 3 decades. However, if scientists would have understood the "homo socialis" with other-regarding preferences already before, the key concept of the "homo economicus" should have disappeared from the economic literature since a long time, but it didn't for a reason. In fact, the increasing empirical and experimental evidence for fairness preferences and unexpectedly high levels of cooperation in one-shot prisoner's dilemma, dictator and ultimatum games have been waiting for a convincing theoretical explanation until very recently. It is important here to distinguish between other-regarding preferences and cooperative ("altruistic") behavior. Other-regarding preferences means that people intentionally do not maximize their own payoffs, but try to consider and improve the benefits of others. However, most game theoretical work is strictly compatible with the concept of "homo economicus", identifying mechanisms that make it advantageous in one way or another to cooperate. For example, if the "shadow of the future" in repeated prisoner's dilemma interactions is long enough, it creates a higher payoff when people cooperate, and that's why they do it. In other words, some mechanisms such as repeated interactions, punishment, transfer payments, and others change the payoff structure of a prisoner's dilemma game such that there is no dilemma anymore. Martin Nowak has mathematically shown that many such mechanisms can be understood with Hamilton's rule, according to which people cooperate when the benefits of cooperation exceed the costs. Other work shows that cooperation in prisoner's dilemma games may survive if people imitate more successful behavior of neighbors, but if one believes in rational choice, why should people imitate, if they can reach a higher payoff by another behavior? In fact, all such cooperation in spatial prisoner's dilemma games disappears, if imitation is replaced by a "best response" rule, which assumes a strict maximization of utility, based on the previous decision of the interaction partners. In Ref. [4], Grund *et al.* have combined such a "best response" rule with standard evolutionary rules of mutation and selection, when people reproduce. The unexpected outcome was a "homo socialis", if offspring stay close to their parents, which they often do. But the transition does not happen automatically by itself. It requires the population to go through a phase where unconditionally "friendly" behavior is dysfunctional, which happens only by "mistake" (due to mutations). For other-regarding preferences to emerge, random spatio-temporal coincidence of people with friendly traits is also important. Remarkably, however,

summarized by the statement "evolution will punish you, if you're selfish and mean" [5]. Is this really true? And what implications would this have for our economic theory and institutions?

In fact, the success of the human species as compared to others results mainly from its social nature. There is much evidence that evolution has created different incentive systems, not just one: besides the desire to possess (in order to survive in times of crises), this includes sexual satisfaction (to ensure reproduction), curiosity and creativity (to explore opportunities and risks), emotional satisfaction (based on empathy), and social recognition (reputation, power). Already Adam Smith noted: "How ever selfish man may be supposed, there are evidently some principles in his nature, which interest him in the fortune of others, and render their happiness necessary to him, though he derives nothing from it."[2]

The social nature of man has dramatic implications, both for economic theory and for the way we need to organize our economy. As we are more and more connected with others, the "homo economicus", i.e. the independent decision-maker and perfect egoist, is no longer an adequate representation or good approximation of human decision-makers. Reality has changed. We are applying an outdated theory, and that's what makes economic crises more severe.

**Outdated theory, outdated institutions**

In fact, recent experimental results suggest that the majority of decision-makers are of the type of a "homo socials" with equity- or equality-oriented fairness preferences [1,6]. The "homo socialis" is characterized by two features: interdependent decision-making, which takes into account the impact on others, and conditional cooperativeness. Note that the "homo socialis" takes self-determined, free decisions. However, in contrast to what we have today, the principle is not to encourage everyone to rip off others, and then afterwards to give back some of the individual benefits to those in need through taxes or philanthropy - as we know, this cannot overcome "tragedies of the commons" (see below).

The "homo socialis" rather decides in a smarter way than the "homo economicus", recognizing that friendly and fair behavior can generate better outcomes for everybody than if everyone or every company is just thinking for himself. Interestingly, putting oneself into the shoes of others when taking decisions creates interdependent decisions, "networked minds". Such "networked minds" enable collective intelligence, i.e. they can take more

---

conditionally cooperative behavior resulting from other-regarding preferences may occur between strangers, i.e. they do not require genetic relatedness, as the following movie shows: http://vimeo.com/65376719. In any case, spatio-temporal correlations (here: the co-evolution of individual preferences and behavior) can promote cooperation more than expected for a payoff-maximizing "homo economicus". **These new discoveries mean that key concepts of economics must be reconsidered.**

[2] Smith, A., The Theory of Moral Sentiments (A. Millar, London, 1759).
[3] This has many things in common with how Swiss-style basic democracy works.

intelligent and better decisions than a single mind can do.

However, other-regarding preferences are vulnerable to exploitation by the "homo economicus". In a selfish environment, the "homo socialis" cannot thrive. In other words, if the settings are not right, the "homo socialis" behaves the same as the "homo economicus". That's probably why we haven't noticed its existence for a long time. Our theories and institutions were tailored to the "homo economicus", not to the "homo socialis".

In fact, some of today's institutions, such as homogeneous markets with anonymous exchange, undermine cooperation in social dilemma situations, i.e. situations in which cooperation would be favorable for everyone, but non-cooperative behavior promises additional benefits [7, Fig. 2].

**New institutions for a global information society**

In the past we have built public roads, parks and museums, schools, libraries, universities, and global markets. What would be suitable institutions for the 21st century? Reputation systems can transfer the success principles of social communities to our globalized society, the global village. Most people and companies care about reputation. Therefore, reputation systems could support socially oriented decision-making and cooperation, with better outcomes for everyone [8]. In fact, reputation systems spread on the Web 2.0 like wildfire. People rate products, sellers, news, everything, be it at amazon, ebay, or trip adviser. We have become a "like it" generation, because we listen to what our friends like.

Importantly, recommender systems should not narrow down socio-diversity, as this is the basis of happiness, innovation and societal resilience. We don't want to live in a filter bubble, where we don't get a good picture of the world anymore, as Eli Pariser has pointed out [9]. Therefore, reputation systems should be pluralistic, open, and user-centric. Pluralistic reputation systems are oriented at the values and quality criteria of individuals rather than recommending what a company's reputation filter thinks is best. Self-determination of the user is central. We must be able to use different filters, choose the filters ourselves, and modify them. The diverse filters would mine the ratings and comments that people leave on the Web, but also consider how much one trusts in certain information sources.

Reputation creates benefits for buyers and sellers. A recent study shows that good reputation allows sellers to take a higher price, while customers can expect a better service [10]. Reputation systems may also promote better quality as well as socially and environmentally friendly production. This could be a new approach to reach more sustainable production, based on self-regulation rather than enforcement by laws. One day, reputation systems may also be used to create a new kind of money. The value of "qualified money" would depend on it's reputation and thereby create incentives to invest in ways that increase a money unit's reputation. It might create a more adaptive financial system and help to mitigate the recurrent crises we are facing since hundreds of years. But the details still have to be worked out.

**Benefits of a self-regulating economy**

Reputation systems could overcome some of the unwanted side effects of anonymous exchange thanks to pseudonymous or personal interactions. Thereby, they could potentially counter "tragedies of the commons" such as global warming, environmental exploitation and degradation, overfishing, ... - constituting some of our major unsolved global problems. We can witness such kinds of "social dilemma problems" everywhere. So far, governments try to fix them with top-down regulations and punitive institutions. However, these are very expensive, and often quite ineffective. Basically all industrialized countries suffer from exploding debts. In many countries, we cannot pay for this much longer, we are at the limit. We need a new approach. As Albert Einstein pointed out: "We cannot solve our problems with the same kind of thinking that created them."

Institutions supporting the "homo socialis" such as suitably designed reputation systems would enable a self-regulation of socio-economic systems. But self-regulation does not mean that everyone can choose the rules he likes. It only works with an other-regarding element. The self-regulation rules must be able to achieve a balance between the interests of everyone, who is affected by the externalities of a decision.

Other-regarding decisions can overcome the classical conflict between economic and social motives. Self-regulation could also overcome the struggle between the bottom-up organization of markets and the top-down regulation by politics. This would remove a lot of friction from our current system, making it much more efficient - in the same way as the transition from centrally planned economies to self-organized markets has often created huge efficiency gains.

This can be illustrated with an example from urban traffic management. Traffic control is a problem where not everybody's desires can be satisfied immediately and at the same time, like in economic systems. It is a so-called NP-hard optimization problem - the computational effort explodes with system size, as for many economic optimization problems, e.g. in production and logistics. The study compares three kinds of control: A centralized top-down regulation by a traffic center, the classical control approach, and two decentralized control approaches. The first one assumes that each intersection independently minimizes the waiting times of approaching vehicles, as a "homo economicus" would do. The second one decides in an other-regarding way: it interrupts the minimization of waiting times, when this is needed to avoid spill-over effects at neighboring intersections. Summarizing, the "homo economicus" approach works well up to a moderate utilization of intersections, but queue lengths get out of control long before the intersection capacity is reached. The bottom-up self-regulation based on the principle of the "homo socialis" approach beats both, the centralized top-down regulation and the bottom-up self-organization based on principles of the "homo economicus". Other-regarding behavior improves the coordination among neighboring intersections. It makes the principle of the "invisible hand" work, even at high utilizations.

| Market System | Centrally Planned Economy | Conventional Market Economy | Participatory Market Society |
|---|---|---|---|
| Agent | Central Planner | Homo Economicus | Homo Socialis |
| Decision-Making | Elite decides for everyone else | Everyone decides himself, but not many have influence | Everyone decides considering others, has influence |
| Organization | Top down | Bottom up | Bottom up |
| Social Structure | Central government, people have no power | Hierarchy of power, representative decisions | Participatory decision-making |
| Regulation | Top down | Top down | Bottom up |
| Wealth Distribution | Flat, equality-oriented distribution | Heavily skewed, social benefit system (redistribution) | Higher average wealth; fair, merit-based distribution |

**Economics 2.0: Emergence of a participatory market society**

But will such a self-regulating system ever be implemented? In fact, this new, third kind of economy is already on its way. The Web 2.0, in particular reputation systems and social media are driving the transition towards an economy 2.0. We see already a strong trend towards decentralized, local production and personalized products, enabled by 3D printers, app stores, and other technologies.

Such developments will eventually create a participatory market society. "Prosumers", i.e. co-producing consumers, the new "makers" movement, and the sharing economy are some examples illustrating this. Just think of the success of Wikipedia, Open Streetmap or Github. Open Streetmap now provides the most up-to-date maps of the world, thanks to more than 1 million volunteers. This is just the beginning of a new era, where production and public engagement will more and more happen in a bottom up way through fluid "projects", where people can contribute as a leaders ("entrepreneurs") or participants. A new intellectual framework is emerging, and a creative and participatory era is ahead. The paradigm shift towards participatory bottom-up self-regulation[3] may be bigger than the paradigm shift from a geocentric to a heliocentric worldview. If we build the right institutions for the information society of the 21st century, we will finally be able to mitigate some very old problems of humanity. "Tragedies of the commons" are just one of them. After so many centuries, they are still plaguing us, but this needn't be.

**Further Reading:**

**Dirk Helbing** is Professor of Sociology, in particular of Modeling and Simulation, and member of the Computer Science Department at ETH Zurich. He earned a PhD in physics and was Managing Director of the Institute of Transport & Economics at Dresden University of Technology in Germany. He is internationally known for his work on pedestrian crowds, vehicle traffic, and agent-based models of social systems. Furthermore, he coordinates the FuturICT Initiative (http://www.futurict.eu), which focuses on the understanding of techno-socio-economic systems, using Big Data. His work is documented by hundreds of scientific articles, keynote lectures and media reports worldwide. Helbing is elected member of the World Economic Forum's Global Agenda Council on Complex Systems and of the German Academy of Sciences "Leopoldina". He is also Chairman of the Physics of Socio-Economic Systems Division of the German Physical Society and co-founder of ETH Zurich's Risk Center.

---

[3] This has many things in common with how Swiss-style basic democracy works.